\documentclass[prb,aps,twocolumn,floatfix,longbibliography]{revtex4-2}
\usepackage[utf8]{inputenc}
\usepackage{graphicx}
\usepackage{epsfig}
\usepackage{color}
\usepackage[tbtags]{amsmath}
\usepackage{hyperref}
\hypersetup{colorlinks,allcolors=black}
\usepackage{amssymb}
\usepackage{gensymb}
\usepackage{float}
\usepackage{amsmath}
\usepackage{tabularx,graphicx}
\usepackage{epstopdf}
\usepackage{latexsym}
\usepackage{color, colortbl}
\usepackage{bm}
\usepackage{titlesec}
\usepackage{slashed}
\usepackage{multirow}
\newcommand{\be}{\begin{equation}}
\newcommand{\ee}{\end{equation}}
\newcommand{\beq}{\begin{eqnarray}}
\newcommand{\eeq}{\end{eqnarray}}
\newcommand{\ba}{\begin{aligned}}
\newcommand{\ea}{\end{aligned}}

\renewcommand{\phi}{\varphi}
\renewcommand{\epsilon}{\varepsilon}

\def \ba{\begin{align*}}
\def \ea{\end{align*}}

\newcounter{indice}

\begin{document}

\title{Effective low energy Hamiltonians and unconventional Landau level spectrum of monolayer C$_3$N }

\author{Mohsen Shahbazi}
\email{mohsenshahbazi1984@yahoo.com}
\affiliation{Department of Physics, Faculty of Science, University of Zanjan, Zanjan, Iran}

\author{Jamal Davoodi}
\affiliation{Department of Physics, Faculty of Science, University of Zanjan, Zanjan, Iran}

\author{Arash Boochani}
\affiliation{Department of Physics, Kermanshah Branch, Islamic Azad University, Kermanshah, Iran}
\affiliation{Quantum Technological Research Center (QTRC), Science and Research Branch, Islamic Azad University, Tehran, Iran}

\author{Hadi Khanjani}
\affiliation{Department of Physics, University of Tehran, P. O. Box 14395-547, Tehran, Iran}

\author{Andor Kormányos}
\email{andor.kormanyos@ttk.elte.hu}
\affiliation{Department of Physics of Complex Systems,
Eötvös Loránd University, Budapest, Hungary}

\begin{abstract}
 We derive a low-energy effective $\mathbf{k}\cdot\mathbf{p}$ Hamiltonians for monolayer  C$_{3}$N at the $ \Gamma $ and $ M $ points of the Brillouin zone 
 where the band edge in the conduction and valence band can be found. Our analysis of the electronic band symmetries     helps to better understand several results of recent \emph{ab-initio} calculations~\cite{tang2022giant,bonacci2022excitonic} for the optical properties of this material. We also calculate the Landau level spectrum. We find that the Landau level spectrum in the degenerate conduction bands at the $ \Gamma $ point acquires properties that are reminiscent  of the corresponding results in bilayer graphene, but there are important differences as well. Moreover, because of the heavy effective mass, $n$-doped samples may host interesting electron-electron interaction effects. 
\end{abstract}

\maketitle

\section{Introduction}
\label{sec:intro}

Graphene\cite{novoselov2004electric} has received a great deal of attention due to its unique mechanical, electronic, thermal and optoelectronic properties \cite{neto2009electronic,balandin2011thermal,falkovsky2008optical}. However,  having a zero  band gap  limited the applications of graphene in electronic nano-devices and  motivated the search for atomically thin two-dimensional (2D) materials which have a finite band gap. 
This lead to the discovery of, e.g.,  monolayer transition metal dichalcogenides~\cite{Heinz2010,TMDC-Naturenano-review,TMDC-Naturephys-review}, silicene~\cite{vogt2012silicene,kara2012review}, 
phosphorene~\cite{carvalho2016phosphorene,batmunkh2016phosphorene}, germanene~\cite{acun2015germanene}.
In recent years, compounds of carbon-nitrides C$_x$N$_y$ have also  become attractive 2D materials~\cite{li2017carbon,makaremi2018band,bafekry2021two}. 
For example graphitic carbon nitride (g-C$_{3}$N$_{4}$), which is a direct  band gap semiconductor, has  potential  applications in photocatalysis and in solar energy 
conversion due to its strong optical absorption at visible frequencies~\cite{ismael2020review,ong2016graphitic}. 
Another carbon-nitride compound, two-dimensional crystalline C$_{3}$N has also been recently synthesized~\cite{c3n-synthesis,YangWeiGang2017}.  C$_{3}$N is an indirect band gap semiconductor 
with energy gap of $ 0.39 $ eV~\cite{YangWeiGang2017}. Moreover, it has  shown  favorable  properties, such as high mechanical stiffness~\cite{mortazavi2017ultra}
and interesting excitonic effects~\cite{bonacci2022excitonic,tang2022giant}. 
In addition, its thermal conductivity properties have  been investigated~\cite{gao2018first,kumar2017ultralow,mortazavi2017ultra} and it has been predicted that the electronic, optical and thermal  properties of monolayer C$_{3}$N  can  be tuned by strain 
engineering~\cite{Zhao2019, chen2019anisotropic}. 

\begin{figure}[t]
\centering
\includegraphics[width=0.35\textwidth]{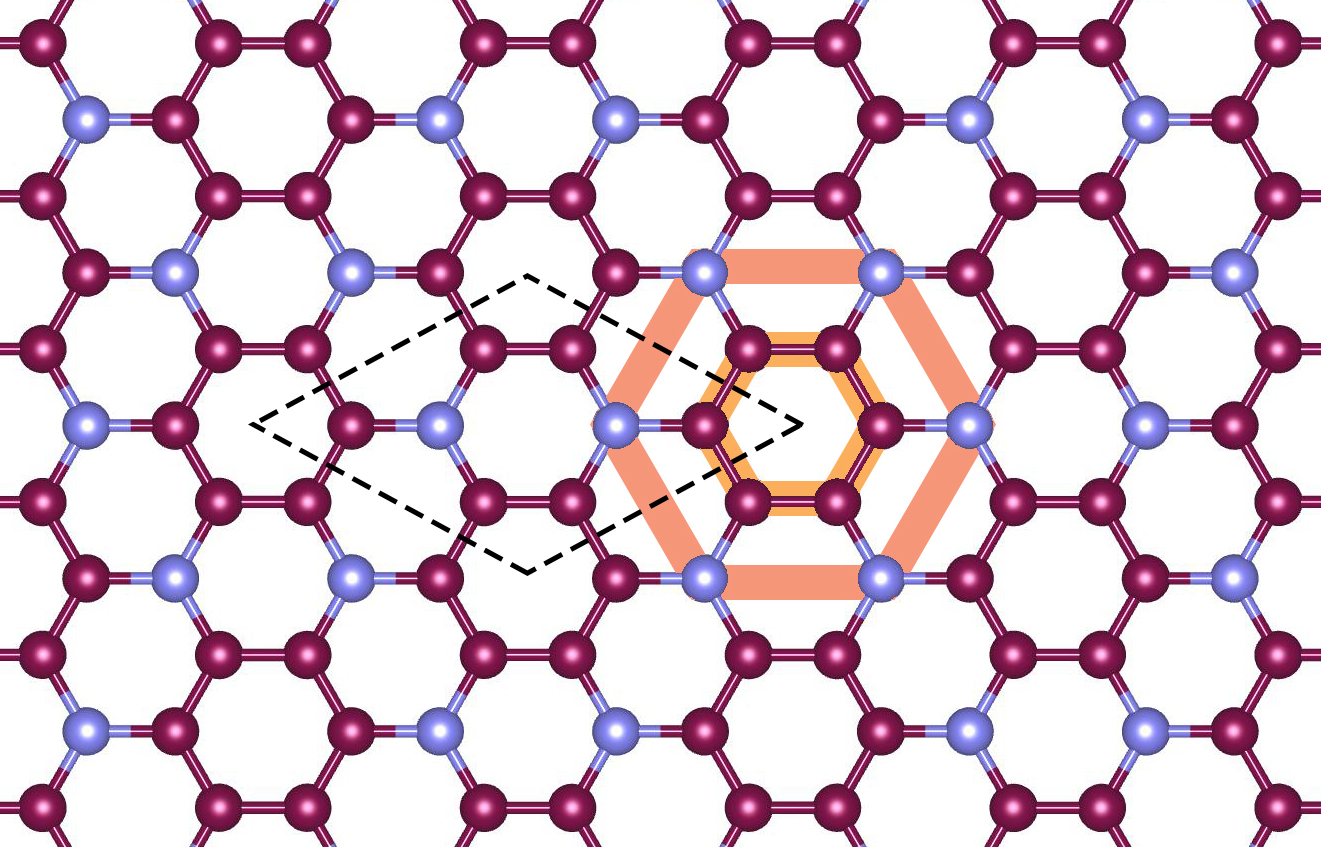}
\caption{a) Crystal structure of C$_{3}$N  monolayer. Purple and blue circles refer to carbon and nitrogen atoms, respectively. Dashed black lines show the unit cell. The orange line show the hexagonal unit cell, which can be useful to understand certain optical properties, see Sec.~\ref{sec:optical}. }
\label{fig:lattice}
\end{figure}
In this work we will employ the  $\mathbf{k}\cdot\mathbf{p}$~\cite{voon2009kp,dresselhaus2007group}  approach in order to  study the electronic properties 
of the monolayer  C$_{3}$N.  We obtain the materials specific parameters appearing in the $\mathbf{k}\cdot\mathbf{p}$ model
from fitting it to density functional theory (DFT)  band structure calculations. A similar methodology   has been successfully used, e.g., 
for monolayers of transition metal dichalcogenides~\cite{kormanyos2015k,kormanyos2013monolayer}. In particular, since the 
conduction band (CB) minimum and valence band (VB) maximum are located at $ \Gamma $ and $ M $ points of the Brillouin zone, respectively, we 
obtain $\mathbf{k}\cdot\mathbf{p}$ Hamiltonians valid in the vicinity of these points.  
The insight given by the $\mathbf{k}\cdot\mathbf{p}$ model allows us to comment on certain optical properties as well. 
Moreover, we will also study the Landau level spectrum of C$_{3}$N, which, to our knowledge, 
has not been considered before.

This paper is organized as follows. In Sec.~\ref{sec:dft-calc} we start with a short recap of the band structure obtained with the help of the 
density functional theory calculations.
In Sec.~\ref{sec:kp-Hamiltonians}, effective $\mathbf{k}\cdot\mathbf{p}$ Hamiltonians at $\Gamma$and  $M$ points are obtained, using symmetry groups and perturbation theory. 
Certain optical properties of this material are discussed in Sec.~\ref{sec:optical}. In Sec.~\ref{sec:LLs} 
the spectra of Landau levels for this material are calculated at the  $\Gamma$ and  $M$ points. Finally, our main results are summarized in Sec.~\ref{sec:summary}.


\section{Band structure calculations}
\label{sec:dft-calc}

The band structure of monolayer C$_3$N has been calculated before at the DFT level of theory~\cite{zhou2017computational,Wang2018,Zhao2019} and also using 
the GW approach\cite{Wu2018,bonacci2022excitonic,tang2022giant}. The main effect of the GW approach is to enhance the band gap and 
this does not affect our main conclusions below. To be self-contained, we repeat the band structure calculations at the DFT level. 
The schematics of the crystal lattice  of single-layer C$_{3}$N  is shown in the Fig.~\ref{fig:lattice}. 
The lattice of  C$_{3}$N  possess P6/mmm space group with a planar hexagonal lattice and the unit cell  contains  six carbon and two nitrogen atoms. 
We used the Wien2K package\cite{SCHWARZBlaha2003259} to perform first-principles calculations based on density functional theory (DFT). 
For the exchange-correlation potential we used  the generalized gradient approximation (GGA)~\cite{perdewRuzsinszky2008}. 
The optimized input parameters such as RKmax, lmax, and k-point were selected to be $8.5$, $10$, and $14\times14\times3$, respectively. 
The convergence accuracy of self-consistent calculations for the electron charge up to $ 0.0001 $ was chosen and the forces 
acting on the atoms were optimized to $0.1  $dyn/a.u.
\begin{figure}
\centering
\includegraphics[width=0.5\textwidth]{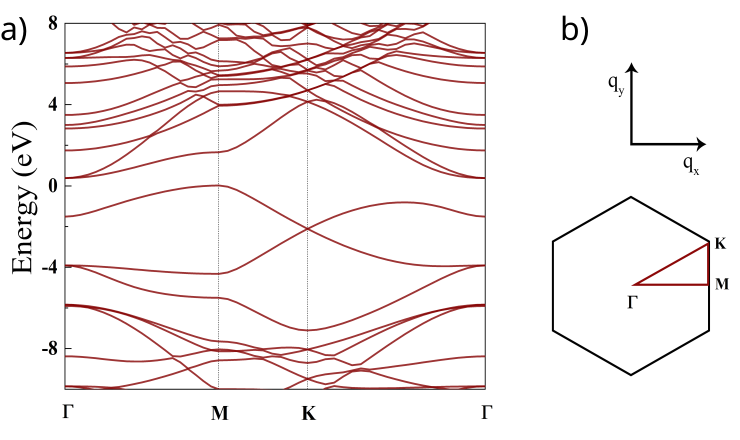}
\caption{a) DFT band structure calculations for C$_{3}$N along the $\Gamma-K-M-\Gamma$ line in the BZ. b) orientation of the BZ and the 
high symmetry points $\Gamma$, $K$, $M$. 
}
\label{fig:bandstructure}
\end{figure}
The optimized lattice constant is $a_0=4.86\AA{}$,  in good agreement with previous studies~\cite{zhou2017computational,bafekry2019c3n}.

The calculated band structure is shown in Fig.~\ref{fig:bandstructure}.  The conduction band minimum  is located at the $\Gamma$ point, while the valence 
band maximum can be found  at the $ M $ of the BZ. Thus, at the DFT level C$_{3}$N is an indirect band gap semiconductor with a band gap of  
 $E_{bg}=0.48$\,eV which is in good agreement with  previous works~\cite{YangWeiGang2017,makaremi2017adsorption}. 
 We have checked that the magnitude of the spin-orbit coupling is  small at the band-edge points of interest and therefore in the following 
 we will neglect it.  The main effect  of spin-orbit coupling is  to
lift degeneracies at certain high-symmetry points and lines, e.g., the four-fold degeneracy of the conduction band at the $\Gamma$ point would 
be split into two, two-fold degenerate bands.


\section{Effective $\mathbf{k}\cdot\mathbf{p}$  Hamiltonians}
\label{sec:kp-Hamiltonians}

We now introduce the $\mathbf{k}\cdot\mathbf{p}$ for the $\Gamma$ point, where the band edge of the CB is located, and for the $M$ point, where the 
band edge  of the VB can be found.

\subsection{$\Gamma $ point}
\label{sec:kp-Gamma}

The pertinent point group at the  $\Gamma $  point of the BZ is $D_{6h}$. We obtained the corresponding irreducible representations of the nine 
bands around the Fermi level at the $\Gamma $  point  with the help of the Wien2k package.  Using  this information one can 
then set up a nine bands $\mathbf{k}\cdot\mathbf{p}$ model along the lines of Ref.~\cite{kormanyos2013monolayer}, see Appendix~\ref{app:Gamma-nine-band} for details. 
Here we only mention that there is no $\mathbf{k}\cdot\mathbf{p}$ matrix element between the VB and the degenerate CB, CB+1 which means that direct optical transitions 
are not allowed between these two bands.
Since it is usually difficult to work with a nine-band Hamiltonian, we derived an effective  low-energy Hamiltonian which 
describes  the two (degenerate) conduction bands and the valence band.
Using the L\"owdin partitioning technique~\cite{lowdin1951note,winkler2003spin} we find that 
\begin{subequations}
\begin{align}
H_{\rm eff}^{\Gamma}=H_{0}^{\Gamma} + H_{\mathbf{k}\cdot\mathbf{p}}^{\Gamma},
\end{align}
\begin{align}
H_{0}^{\Gamma}=\begin{pmatrix}
\varepsilon_{vb}&0&0\\
0&\varepsilon_{cb}&0\\
0&0&\varepsilon_{cb+1}\\
\end{pmatrix}
\label{eq:H0-G}
\end{align}
\begin{align}
H_{\mathbf{k}\cdot\mathbf{p}}^{\Gamma}=\begin{pmatrix}
\alpha_{1}q^{2}&0&0\\
0 & (\alpha_{2}+\alpha_3)q^{2}& -\alpha_{3}(q_{+})^{2}\\
0 &-\alpha_{3} (q_{-})^{2} & (\alpha_{2}+\alpha_3) q^{2}
\end{pmatrix}.
\label{eq:kp-G}
\end{align}
\label{eq:Heff-G}
\end{subequations}
Here  $\varepsilon_{cb}=\varepsilon_{cb+1}=0.386$\,eV and $\varepsilon_{vb}=-1.50$\,eV are band edge energies of the degenerate CB minimum 
and VB maximum. 
The wavenumbers $q_x$, $q_y$ are measured from the $\Gamma$ point, $q_{\pm}=q_{x}\pm iq_{y} $ and $q^2=q_x^2+q_y^2$ and in $\alpha_2$ we took into account 
the free electron term\cite{kormanyos2015k}. 

Note, that there are no linear-in-$\mathbf{q}$ matrix elements between the VB and the degenerate CB, CB+1 bands. 
In higher order of $\mathbf{q}$ these bands do couple, but this is neglected in the minimal model given in  Eq.~(\ref{eq:kp-G}). 
The minimal model given in Eq.~(\ref{eq:Heff-G}) already captures an important property of the degenerate CB and CB+1 bands from the DFT calculations, 
which is that their effective masses are different. 
One finds from Eq.(\ref{eq:kp-G}) that the effective masses  are 
$1/m_{cb}^{\Gamma}=\frac{2}{\hbar^2}\alpha_2$   and $1/m_{cb+1}^{\Gamma}=\frac{2}{\hbar^2}(\alpha_2+2\alpha_3)$.
The material parameters $\alpha_i$ can be obtained, e.g., by fitting  these effective masses  to the DFT band structure calculations. 
We found that $\alpha_{1} = 13.77$\,eV\AA$^{2}$, $\alpha_{2} = 5.19$\,eV\AA$^{2}$ and $ \alpha_{3} = 3.95$\,eV\AA$^{2}$. 
The corresponding  effective masses at the $ \Gamma $ point are shown in Table~\ref{tbl:eff-mass}.
\begin{table}[htb]
\centering
\caption{Effective masses at the $ \Gamma $ and M points }
\begin{tabular}{c |c | c|c}\hline
 &All directions& M - $\Gamma$  line&  M - K line  \\\hline
$m_{vb}^{ \Gamma }/m_{e}$ &0.27 &-&- \\\hline
$m_{cb}^{\Gamma}/m_{e}$ & 0.73 &-&- \\\hline
$m_{cb+1}^{\Gamma}/m_{e}$ &0.29&-&-\\\hline
$m_{vb}^{M}/m_{e}$&-&-0.82&-0.12\\\hline
$m_{cb}^{M}/m_{e}$&-&-0.87&0.10
\end{tabular}
\label{tbl:eff-mass}
\end{table}


\subsection{\textbf{M}  point}
\label{sec:kp-M}

Next we consider the $ M $ point, where the location of the VB  maximum is. The relevant point group is  $ D_{2h} $. 
Since this point group has only one-dimensional irreducible representation,  one expects  that there are no  degenerate bands  near the $M$ point. 
This is in agreement with  our DFT calculations, see Fig.~\ref{fig:bandstructure}. Because of the dense spectrum in the conduction band, we 
start with a $13$ bands $\mathbf{k}\cdot\mathbf{p}$ Hamiltonian (see Appendix~\ref{app:M-thirteen-band} for details) and by projecting out the higher energy bands we 
obtain an effective two band model for the VB and the CB. 

This effective model reads
\begin{subequations}
\begin{equation}
H_{\rm eff}^{M}=H_{0}^{M} + H_{\mathbf{k}\cdot\mathbf{p}}^{M},
\end{equation}
\begin{equation}
H_{0}^{M}=\begin{pmatrix}
\varepsilon_{vb}&0\\
0&\varepsilon_{cb}\\
\end{pmatrix},
\label{eq:H0-M}
\end{equation}
 \begin{equation}
 H_{\mathbf{k}\cdot\mathbf{p}}^{M}=\begin{pmatrix}
 \dfrac{\hslash^{2}}{2m_{e}} q_{x}^{2}+\beta_{1} q_{y}^{2}&\gamma_{21}q_{x}\\
 \gamma_{21}^{*}q_{x}&\beta_{2} q_{x}^{2}+\beta_{3} q_{y}^{2}\\
 \end{pmatrix}.
\label{eq:kp-M}
\end{equation}
\label{eq:Heff-M}
\end{subequations}
Here  $ \varepsilon_{vb} = 0.021 $\,eV and $ \varepsilon_{cb} = 1.65$\,eV  refer to  band edges energies of  the VB  and the CB, respectively and the $q_x$ direction is along the 
$\Gamma-M$ line.
We note that   $ H_{\mathbf{k}\cdot\mathbf{p}}^{M} $ includes free electron term as well\cite{kormanyos2015k}. 
It is interesting  to note that  $ H_{\mathbf{k}\cdot\mathbf{p}}^{M} $ has the same general form as the  $\mathbf{k}\cdot\mathbf{p}$ model for the $\Gamma$ 
point of phosphorene~\cite{Zhou2015,Pereire-LLs}. An important difference between the two cases, apart from the fact that the multiplicity 
of the $\Gamma$ and $M$ points are different, is that in the case of C$_3$N there is a saddle point in the dispersion at the $M$ points, 
whereas  in the case of  phosphorene the dispersion  has a positive slope in every direction  at the $\Gamma$ point. 
The  material parameters  appearing Eq.~(\ref{eq:kp-M})  can be obtained  by fitting the dispersion  to the DFT band structure calculations. The range of fitting was $\%0.5$ of both the $M-K$ and $M-\Gamma$ directions.
We found $ \beta_{1}=-31.36$\,eV\AA$^{2}$,   
$ \beta_{2}=-10.02 $\,eV\AA$^{2}$, 
$ \beta_{3}=34.87 $\,eV\AA$^{2}$, and 
$ \gamma_{21}=3.41 $\,eV\AA.
The corresponding effective masses are given in Table~\ref{tbl:eff-mass}. One can indeed see that the effective masses have a different sign along the 
$M-K$ and $M-\Gamma$ directions.


\section{Comments on the optical properties}
\label{sec:optical}

Recently,  Ref.~\cite{tang2022giant,bonacci2022excitonic} have studied the optical properties of C$_3$N based on the DFT+G$_0$W$_0$ methodology to obtain an 
improved value for the band gap and the Bethe-Salpeter approach to calculate the excitonic properties. 
 Several findings of Ref.~\cite{tang2022giant,bonacci2022excitonic} can be interpreted with the help of the results presented in Sec.~\ref{sec:kp-Hamiltonians}. 

According to the  numerical calculations of  Ref.~\cite{bonacci2022excitonic},  the lowest energy direct excitonic state is doubly-degenerate and  dark.  
The corresponding electron-hole transitions are located  in the vicinity of the $\Gamma$ point and the 
electron part of the exciton wavefunction is localized on the benzene rings of C$_3$N  if the hole is fixed on an N atom. 
\begin{figure*}[htb]
 \includegraphics[width=0.5\textwidth]{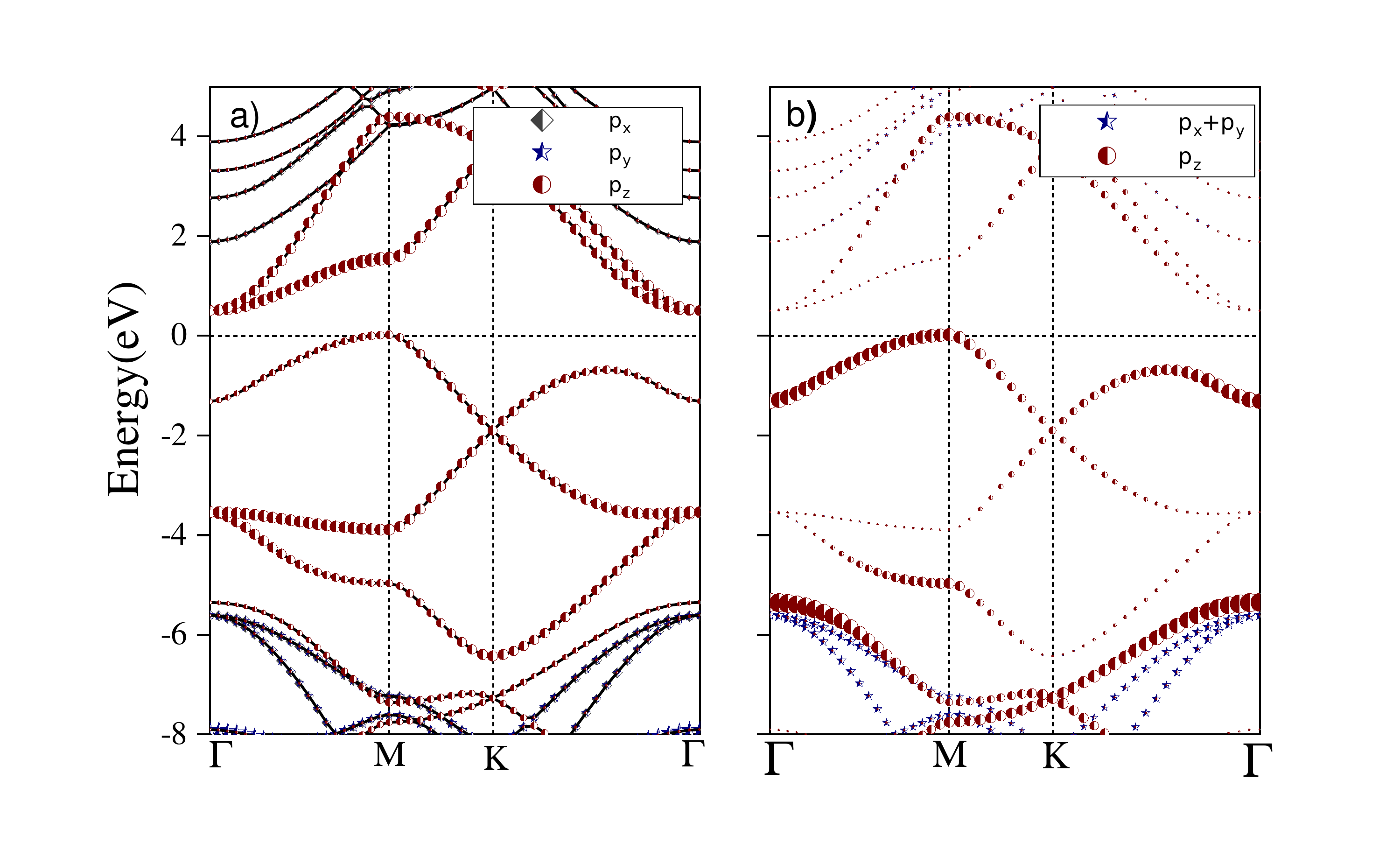}
 \caption{Atomic orbital weight of a) carbon and b) nitrogen atoms in the energy bands of C$_{3}$N monolayer. }
 \label{fig:orbitals}
 \end{figure*}
Firstly, we note  that there is no  $\mathbf{k}\cdot\mathbf{p}$ matrix element between the VB and the doubly degenerate CB at the $\Gamma$ point 
(see Appendix~\ref{app:Gamma-nine-band}) which indicates that direct optical transitions are forbidden by symmetry. 
Furthermore, as shown in Fig.~\ref{fig:orbitals}, 
at the $\Gamma $ point  the $p_z$ atomic orbitals of the C atoms have a large weight in the CB, CB+1 bands and the same applies to
the $p_z$ atomic orbitals of the N atoms in the VB.
According to Table~\ref{tbl:G-bands-irreps}, the degenerate CB, CB+1 bands correspond to the two-dimensional $E_{2u}$ irreducible representation of $D_{6h}$. 
One can  check that the following linear combinations of the  $p_z$ atomic orbitals of the C atoms transform as the partners of the 
$E_{2u}$ irreducible representation:
\begin{subequations}
\begin{align}
&\phi_{1}  = \frac{1}{\sqrt{6}}\left[ p^{(1,C)}_z+ \omega p^{(2,C)}_{z}+ \omega^{2} p^{(3,C)}_{z}\right. \nonumber\\
&\qquad+ \left. p^{(4,C)}_{z}+ \omega p^{(5,C)}_{z}+ \omega^{2} p^{(6,C)}_{z}\right], 
\label{eq:phi1}
\end{align}
\begin{align}
&\phi_{2} = \frac{1}{\sqrt{6}}\left[p^{(1,C)}_z + \omega^{2} p^{(2,C)}_z + \omega p^{(3,C)}_z\right. \nonumber\\
&\qquad + \left.p^{(4,C)}_z + \omega^{2} p^{(5,C)}_z+ \omega p^{(6,C)}_z\right], 
\label{eq:phi2}
\end{align}
\label{eq:phi12}
\end{subequations}
where $ \omega=e^{2\pi i/3} $, and   $p^{(i,C)}_z$, $ i=1,2,...,6 $ denotes the $p_z$ orbitals of the six carbon  
atoms in the unit cell,  see Fig.~\ref{fig:lattice}. Bloch wavefunctions based on $\phi_1$ and  $\phi_2$ 
would indeed have a large weight on the benzene rings in each unit cell, and this helps to explain the corresponding finding 
of  Ref.~\cite{bonacci2022excitonic}. 

On the other hand, the $\mathbf{k}\cdot\mathbf{p}$ matrix elements are non-zero between pairs of the  degenerate VB-1, VB-2 and CB, CB+1 bands, 
see Table~\ref{tbl:kp-nine-band-G}. This means that optical transitions are allowed by symmetry and if circularly polarized light is used 
for excitation then $\sigma^{+}$ and $\sigma^{-}$ would excite transitions between different pairs of bands. Some of the higher energy 
bright excitonic states found in Ref.~\cite{bonacci2022excitonic} should correspond to this transition. 

At the $M$ point, on the other hand, there is finite matrix element between the VB and the CB, see Eq.~(\ref{eq:kp-M}), which suggests 
that optical transitions are allowed by symmetry along the $\Gamma-M$ line. Moreover, one can expect that the optical density of states is 
large going from $M$ towards $\Gamma$ because the VB and the CB are approximately parallel. Note that the time reversed states can be 
found at $-q_x$, i.e., on the other side of the $\Gamma$ point. Therefore  one can expect that in zero magnetic field two, 
degenerate bright excitonic states can be excited,  and in  $\mathbf{k}$ space they  are localized on opposite sides of the $\Gamma$ 
point along the $\Gamma-M$ lines. This correspond to the  findings of  Ref.~\cite{bonacci2022excitonic}. 
In external magnetic field, which breaks time reversal symmetry, the degeneracy  of the two excitonic states  would be broken. 
This is reminiscent of the valley degeneracy 
breaking for magnetoexcitons in monolayer TMDCs~\cite{MagnetoPL-paper,Heinz2014,Atac2015,XiaodongXu2015,Urbaszek2015}.

The  transition along the $\Gamma-M$ line can be excited by linearly polarized light. For a general direction of the linear polarization 
with respect to the crystal lattice, transitions along all three $\Gamma-M$ directions in the BZ would be excited. 
However, when the polarization of the light is perpendicular to one of the $\Gamma-M$ line, then the interband transitions 
are excited only in the remaining two  $\Gamma-M$ ``valleys''. In contrast, when circularly polarized light is used 
for excitation, all three $\Gamma-M$ ``valleys'' are excited.


\section{Landau levels} 
\label{sec:LLs}

We now consider the Landau levels spectrum of monolayer C$_3$N. Using the $\mathbf{k}\cdot\mathbf{p}$ Hamiltonians, one can employ  the Kohn–Luttinger prescription, 
see, e.g., Ref.~\cite{winkler2003spin}. 
This means that one can replace the wavenumber $ \mathbf{q}=(q_{x},q_{y})$ by the operator 
$\mathbf{\hat{q}} = 
\frac{1}{i} \nabla +\frac{e}{\hbar}\mathbf{A}
$,
where $ e>0 $ is the magnitude of electron charge and $\mathbf{A}$ is the vector potential describing the magnetic field. 
In the following we will use the  Landau gauge and $ \mathbf{A}=(0,B_{z}x,0)^{T} $. Since the components of $\mathbf{\hat{q}}$ do not commute, the 
 Kohn–Luttinger prescription should be performed in the original nine-band ($\Gamma$ point) or thirteen-band ($M$ point) 
 $\mathbf{k}\cdot\mathbf{p}$ Hamiltonians (see Appendices \ref{app:Gamma-nine-band}, \ref{app:M-thirteen-band})
 and not in the low energy effective ones given in Eq.~(\ref{eq:Heff-G}) and Eq.~(\ref{eq:Heff-M}), respectively.  
 After the Kohn–Luttinger prescription in the higher dimensional $\mathbf{k}\cdot\mathbf{p}$ model 
 one can again use the Löwdin-partitioning to obtain low-energy effective Hamiltonians by taking care of the order of the non-commuting operators appearing 
 in the downfolding procedure. This approach  was used, e.g., in the case of monolayer TMDCs to study the valley-degeneracy breaking 
 effect of the magnetic field~\cite{kormanyos2015landau}.


\subsection{Effective model at the $\Gamma$ point }

The low-energy model can be expressed in terms of $\hat{q}_{\pm}=\hat{q}_{x}\pm i\hat{q}_{y}$. 
Note, that $ \hat{q}_{+} $ and $\hat{q}_{-} $ do not  commute, so that $ [\hat{q}_{-},\hat{q}_{+}]=\frac{2e\mathit{B}_{z}}{\hslash} $, 
where $ \mathit{B}_{z} $ is a perpendicular magnetic field.  
\begin{align}
\mathcal{H}_{\rm eff}^{\Gamma}=\mathcal{H}_{0}^{\Gamma} + \mathcal{H}^{\Gamma}(B_z)+\mathcal{H}_Z,
\end{align}
where $ \mathcal{H}_{0}^{\Gamma} $ was defined in Eq.~(\ref{eq:Heff-G}), $\mathcal{H}_Z=\frac{1}{2}g_{e}\mu_{b}\mathit{B}_{z}\mathit{S}_{z}$ is the Zeeman term, and  
\begin{align}\label{}
\mathcal{H}^{\Gamma}(B_z)=\begin{pmatrix}
\hat{\mathit{h}}_{11} & 0 & 0 \\ 
 0 & \hat{\mathit{h}}_{22} & \hat{\mathit{h}}_{23}\\
0 & \hat{\mathit{h}}_{32} & \hat{\mathit{h}}_{33}\\
\end{pmatrix}.
\label{eq:kp-G-Bfield}
\end{align}
The  operators $\hat{h}_{ij}$ in $\mathcal{H}^{\Gamma}(B_z)$ are defined as follows: 
\begin{equation}
 \hat{\mathit{h}}_{11} =\frac{1}{2 m_{vb}^{\Gamma}}\left(\hat{q}_+\hat{q}_-+e\hbar B_z\right). 
\end{equation}
\begin{subequations}
\begin{align}
\hat{\mathit{h}}_{22}=(\alpha_2+\alpha_3)\hat{q}_+\hat{q}_{-}+\alpha_2\frac{2 e B_z}{\hbar}-\frac{\hbar e B_z}{2 m_e} 
\end{align}
\begin{align}
\hat{\mathit{h}}_{23}=-\alpha_3^{}(\hat{q}_{+})^{2}, \hspace{0.5cm} \hat{\mathit{h}}_{32}=-\alpha_3^{}(\hat{q}_{-})^{2}, 
\end{align}
\begin{align}
\hat{\mathit{h}}_{33}=(\alpha_2+\alpha_3)\hat{q}_+\hat{q}_{-}+\alpha_3\frac{2 e B_z}{\hbar}+\frac{\hbar e B_z}{2 m_e}.
\end{align}
\label{eq:hs-G-Bfield}
\end{subequations}
Here $\hat{h}_{11}$ corresponds to the VB, while the degenerate CB, CB+1 are described by the $2\times 2$ block in Eq.~(\ref{eq:kp-G-Bfield}).
One can introduce the creation  and annihilation  operators $a^{\dagger}$, $a$ by $\hat{q}_{-}=\frac{\sqrt{2}}{l_{B}} a $, 
$ \hat{q}_{+}=\frac{\sqrt{2}}{l_{B}} a^{\dagger}$, where $ l_{B}=\sqrt{\hslash/eB_{z}} $  is the magnetic length. 
The LLs are obtained  from $\hat{\mathit{h}}_{11} $ correspond to the usual harmonic oscillator spectrum $E_{vb}^{\Gamma}=\hbar\omega_{vb}^{\Gamma}(n+1/2)$, where 
$\omega_{vb}=eB_z/m_{vb}^{\Gamma}$ and $n=0,1,2\dots$ is a positive integer.

Regarding the LLs of the degenerate CB, CB+1 bands, one can anticipate from Eq.~(\ref{eq:hs-G-Bfield}) that the Landau level spectrum of 
this minimal model exhibits an interplay of features known from bilayer graphene and conventional semiconductors. 
Two eigenstates read 
\begin{equation}
 \Psi_0=\left(
 \begin{array}{c}
 |0\rangle\\
  0
 \end{array}
\right), \hspace{0.5cm} 
\Psi_1=\left(
 \begin{array}{c}
 |1\rangle\\
  0
 \end{array}
\right),
\label{eq:Psi01}
\end{equation}
where $|0\rangle$ and $|1\rangle$ are harmonic oscillator eigenfunctions. 
The corresponding 
eigenvalues are 
$E_0^{\Gamma}=\alpha_2\frac{2 e B_z}{\hbar}-\frac{\hbar e B_z}{2 m_e}
$ 
and 
$E_1^{\Gamma}=(2\alpha_2+\alpha_3)\frac{2 e B_z}{\hbar}-\frac{\hbar e B_z}{2 m_e}
$, respectively. 
The eigenstates in Eq.~(\ref{eq:Psi01}) have the same form as the two lowest energy eigenstates of bilayer graphene.  However, 
the $E_0$ and $E_1$ are not degenerate and they  do depend on magnetic field, unlike in the case of bilayer graphene. 

The rest of the  eigenvalues can be obtained by using the Ansatz 
\begin{equation}
\Psi_n=\left(
 \begin{array}{c}
 a_1 |n+2\rangle\\
 a_2 |n \rangle
 \end{array}
\right), 
\label{eq:Psi_n}
\end{equation}
where $n=0,1,2\dots$ and $a_1$,$a_2$ are constants. This Ansatz leads to a $2\times 2$ eigenvalue equation yielding the energy of two LLs for 
each $n$. The eigenvalues can be analytically calculated, but the resulting expression is quite lengthy and not particularly insightful. 

\begin{figure}
\centering
\includegraphics[width=0.45\textwidth]{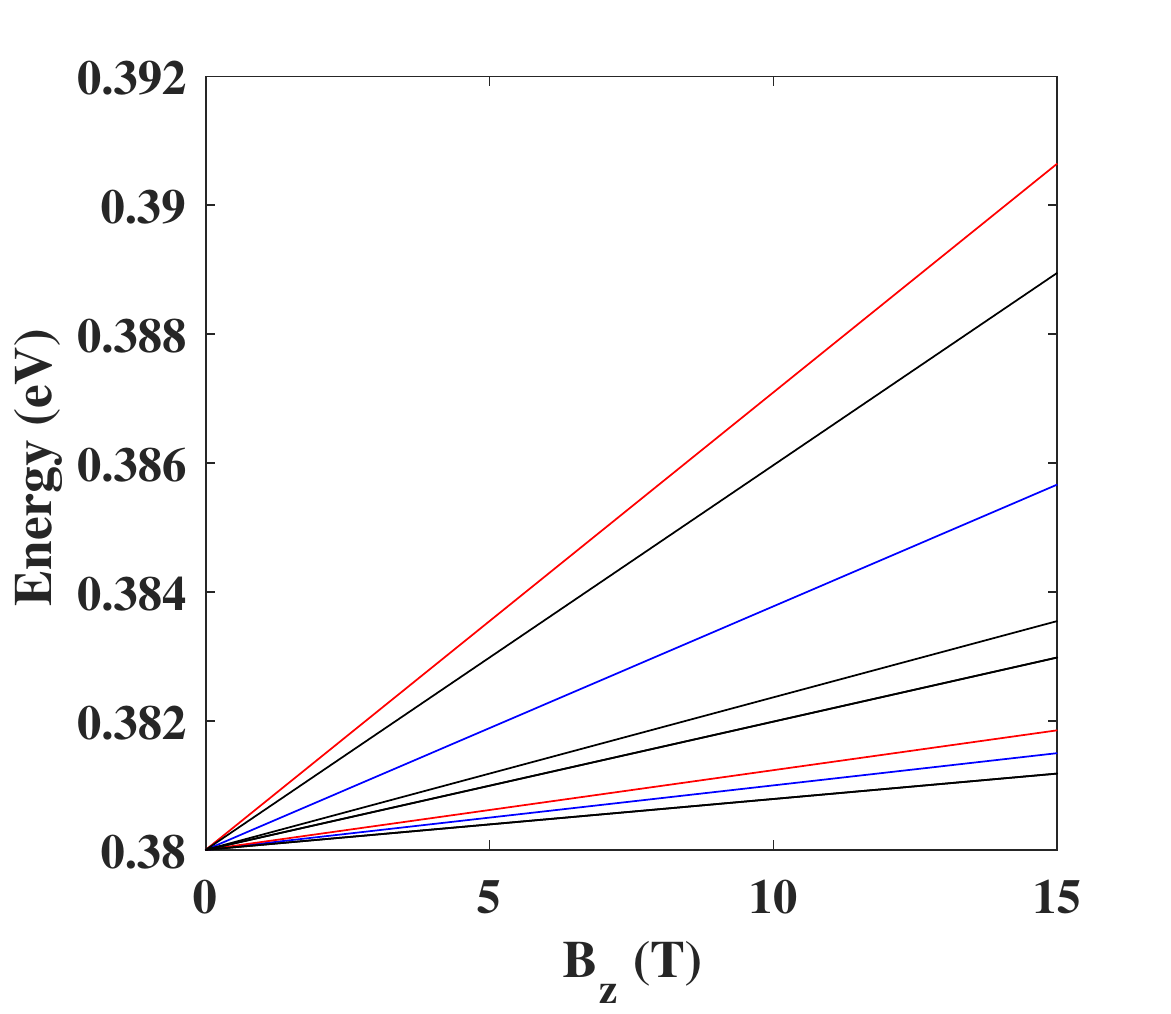}
\caption{Landau levels in the CB at the $\Gamma$ point of the BZ as a function of the out-of-plane magnetic field $B_z$. Blue lines show $E_0^{\Gamma}$ and $E_1^{\Gamma}$ given below Eq.~(\ref{eq:Psi01}) and  red lines indicate the first two LLs that can be obtained from the Ansatz in Eq.~(\ref{eq:Psi_n}). Black lines show the "conventional" LLs, see the text.}
\label{fig:LL-cb-G}
\end{figure}
\begin{figure}
\centering
\includegraphics[width=0.45\textwidth]{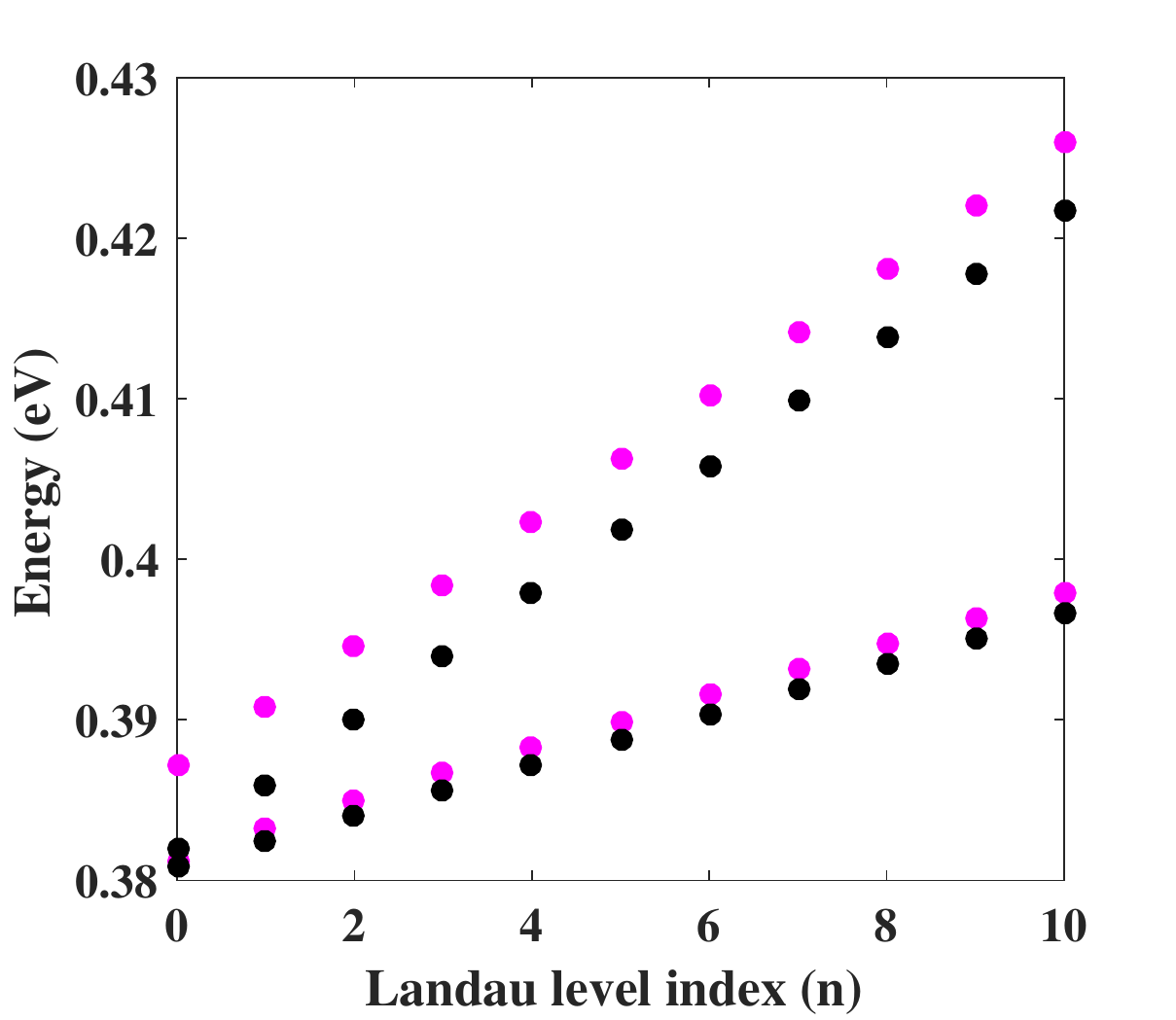}
\caption{Landau levels in the CB at the $\Gamma$ point of the BZ as a function of the LL index n for $B_z=10$ T.   Magenta dots indicate LLs which can be obtained from the Ansatz in Eq.~(\ref{eq:Psi_n}). Black dots show the "conventional" LLs, see the text.}
\label{fig:LL-n-G}
\end{figure}

We plot the first four LLs 
as a function of magnetic field in Fig.~\ref{fig:LL-cb-G}. They correspond to $E_0^{\Gamma}$ and $E_1^{\Gamma}$ given below Eq.~(\ref{eq:Psi01}) and the two LLs that can be obtained using the Ansatz in  Eq.~(\ref{eq:Psi_n}) for $n=0$. 
For comparison, we also plot the energies of "conventional" LLs  $E_n^{cb}=\hbar\omega_{cb}^{\Gamma}(n+\frac{1}2)$ and $E_n^{cb+1}=\hbar\omega_{cb+1}^{\Gamma}(n+\frac{1}2)$, where 
$\omega_{cb}^{\Gamma}={e B_z}/{m_{cb}^{\Gamma}}$ ($\omega_{cb+1}^{\Gamma}={e B_z}/{m_{cb+1}^{\Gamma}}$) and the effective mass $m_{cb}^{\Gamma}$ ($m_{cb+1}^{\Gamma}$) was 
defined below Eq.~(\ref{eq:Heff-G}). One can see that the LLs calculated using  Eqs.~(\ref{eq:Psi01}) and 
(\ref{eq:Psi_n}) are  different from the conventional LLs, which indicates the important effect of the interband coupling. In Fig.~\ref{fig:LL-n-G} we show 
the LL energies as a function of the LL index $n$ for a fixed magnetic field $B_z=10$\,T. One can see that 
for large $n$ the LLs energies obtained from Eq.~\ref{eq:Psi_n} (magenta dots) run parallel with 
the conventional LLs (black dots). This  means  that in this limit both set of LLs can be described by  
cyclotron energies  $\hbar\omega_{cb}^{\Gamma}$ and $\hbar\omega_{cb+1}^{\Gamma}$, but there is an energy difference between them. However, in the deep quantum 
regime ($n=0,1$) the two sets of LLs  cannot be characterized by the same cyclotron frequencies. 

We note, however, that electron-electron interactions may modify the above single particle LL picture, similarly to what has recently been found in 
monolayer MoS$_2$~\cite{ETH-paper}.
Namely, the obtained effective masses of the degenerate CB and CB+1 bands are quite large, which means that the kinetic energy of the electrons is suppressed. 
The importance of the electron-electron interactions can be characterized by the dimensionless Wigner-Seitz radius $r_s=1/(\sqrt{\pi n_e} a_B^{\ast})$. 
Here $n_e$ is the electron density,  
 $a_B^{\ast}= a_B (\kappa m_e / m^{\ast})$ is the effective Bohr radius, $m^{\ast}$ is the effective mass, 
$\kappa$ is the dielectric constant and $a_B$ is the Bohr radius. Taking~\cite{zhou2017computational}  $\kappa=5$ and an 
electron density of $n=4\cdot 10^{12}$\,cm$^2$, one finds $r_s=7.79$  for the heavier band, where $m^{\ast}/m_e=0.73$. 
This value of $r_s$ indicates that electron-electron interactions can be important. Therefore this material
may offer an interesting system to study the interplay of electron-electron interactions and interband coupling. 


\subsection{$M$ point}

We obtain the following low-energy effective Hamiltonian in terms of the operators $\hat{q}_x$ and $\hat{q}_y$:
\begin{align}
\mathcal{H}_{\rm eff}^{M}=\mathcal{H}_{0}^{M}+\mathcal{H}^{M}(B_z)+\mathcal{H}_Z
\end{align}
where $ \mathcal{H}_{0}^{M} $ was defined in Eq.~(\ref{eq:H0-M})  and $\mathcal{H}^{M}(B_z)$ corresponds to Eq.~(\ref{eq:kp-M}):
\begin{align}\label{e123}
\mathcal{H}^{M}(B_z)=\begin{pmatrix}
\hat{\mathit{h}}_{11} & \hat{\mathit{h}}_{12}\\
\hat{\mathit{h}}_{21} & \hat{\mathit{h}}_{22}\\
\end{pmatrix},
\end{align}
 where 
 \begin{subequations}
\begin{align}
\hat{\mathit{h}}_{11}=\dfrac{\hslash^{2}}{2m_{e}}\hat{q}_{x}^{2}+\beta_{1}\hat{q}_{y}^{2}, 
\end{align}
\begin{align}
\hat{\mathit{h}}_{12}=\gamma_{21}\hat{q}_{x}, \hspace{0.5cm} \hat{\mathit{h}}_{21}=\gamma_{21}^{*}\hat{q}_{x},
\end{align}
\begin{align}
\hat{\mathit{h}}_{22}=\beta_{2}\hat{q}_{x}^{2}+\beta_{3}\hat{q}_{y}^{2}. 
\end{align}
\end{subequations}
The off-diagonal elements  of $\mathcal{H}^{M}(B_z)$ are significantly smaller then the diagonal ones 
for magnetic fields $B_z\lesssim 15$\,T, therefore one can again use the L\"owdin 
partitioning to transform them out. Since the band edge at the $M$ point is in the VB, in the following we consider  the LLs in this band. 
By rewriting $\hat{q}_{x}$ and $ \hat{q}_{y}$ in terms of annihilation  and creation operators
$ \hat{q}_{x}=\dfrac{1}{\sqrt{2} l_{B}} (\hat{a}^{\dagger}+\hat{a})$, and 
$ \hat{q}_{y}=\dfrac{-i}{\sqrt{2} l_{B}} (\hat{a}^{\dagger}-\hat{a})$, one finds that the LLs for VB are given by
\begin{align}
{E}_{n,vb}=\varepsilon_{vb} + \hslash \omega_{vb}^{M} (n+\frac{1}{2})
+\dfrac{1}{2}g_{e}\mu_{b}B_{z}S_{z}.
\end{align}
Here $\omega_{vb}^{M}=\dfrac{e\mathit{B}_{z}}{2m_{vb}^{*}}$  is cyclotron frequency, where $m_{vb}^{}=\sqrt{m_{vb,x} m_{vb,y}}$ and 
$m_{vb,x}$ and $m_{vb,y}$ refer to the effective masses along the $\Gamma$-$M$ and $M$-$K$ directions, see Table~\ref{tbl:eff-mass}. 


\section{Summary}
\label{sec:summary}
In summary, we  derived effective low energy Hamiltonians for monolayer C$_{3}$N at the $\Gamma$ and $M$ points of the Brillouin zone. We showed that at the $\Gamma$ point  there are no linear-in-$\mathbf{q}$ matrix elements between the VB and the degenerate CB bands, which means that optical transition are not allowed.  We showed that optical transitions are allowed  between the degenerate VB-1, VB-2 and the CB, CB+1 bands at the $\Gamma$ point if circularly polarized light is used.
We also found that there is a saddle point in the energy dispersion of $M$ point.   In addition, we suggested that the  transition along the $\Gamma-M$ line can be excited by linearly polarized light at the $M$ point. Moreover, we obtained the Landau level spectra  by employing  the Kohn–Luttinger prescription for the $\mathbf{k}\cdot\mathbf{p}$ Hamiltonians at the $\Gamma$ and $M$ points. We pointed out the important effect of the interband coupling on the Landau level spectrum at the $\Gamma$ point. An interesting further direction could be the study of electron-electron interaction effects on the Landau level splittings, which can be important due to the heavy effective mass.


\section{Acknowledgments}
We thank Martin Gmitra for his technical help at the early stages of this project. 
This work was supported by the ÚNKP-21-5 New National Excellence Program of the Ministry for Innovation and Technology 
from the source of the National Research, 
Development and Innovation Fund and by the  Hungarian Scientific Research Fund (OTKA) Grant No. K134437.  
A.K. also acknowledges support from the Hungarian
Academy of Sciences through the Bólyai János Stipendium (BO/00603/20/11).


\appendix

\section{Nine-band model at the $ \Gamma $ point}
\label{app:Gamma-nine-band}

In this Appendix, we provide some details of the nine-band  $\mathbf{k}\cdot\mathbf{p}$ Hamiltonian mentioned in the main text, 
which is valid at  $ \Gamma $ point of the BZ. The small group of the $\mathbf{k}$ vector at the $ \Gamma $ point is a $ D_{6h}$, the 
character table  of this point group is given in  Table~\ref{tbl:D6h}.

\begin{table*}[!htb]
\centering
\caption{Character table for $ D_{6h} $  point group.}
\begin{tabular}{c c c c c c c c c c c c c c c}\hline
$D_{6h}$ &$ E$ & $2C_{6}(z)$ &$2C_{3}$&$C_{2}$ & $3C_{2}^{'}$& $3C_{2}^{"}$&$ i $&$2S_{3}$&  $2S_{6}$ &$\sigma_{h}(xy) $&$ 3\sigma_{d} $&$ 3\sigma_{v}$ &Linear functions, rotations\\\hline
$ A_{1g} $ &  +1 & +1 & +1 & +1 & +1 &+1 &+1 & +1 &  +1 & +1 &+1 &+1&- \\
$A_{2g}$& +1 & +1 & +1 & +1 & -1 & -1 & +1 & +1 &+1&+1&-1&-1&$ R_{z} $ \\
$B_{1g}$& +1 & -1 & +1 & -1 & +1 & -1 & +1 & -1 &+1&-1&+1&-1&-\\
$ B_{2g} $& +1 & -1 & +1 & -1 & -1 & +1 & +1 & -1 &+1&-1&-1&+1&- \\
$E_{1g}$& +2 & +1 & -1 & -2 & 0 & 0 & +2 & +1  & -1 & -2 & 0 & 0& $ (R_{x},R_{y}) $ \\
$ E_{2g} $& +2 & -1 & -1 & +2 & 0 & 0 & +2 & -1 &-1&+2&0&0&- \\
$ A_{1u} $& +1 & +1 & +1 & +1 & +1 & +1 & -1 & -1 &-1&-1&-1&-1&-\\
$ A_{2u} $ & +1 & +1 & +1 & +1 & -1 & -1 & -1 & -1 &-1&-1&+1&+1&z \\
$ B_{1u} $ & +1 & -1 & +1 & -1 & +1 & -1 & -1 & +1 &-1&+1&-1&+1&- \\
$ B_{2u} $ & +1 & -1 & +1 & -1 & +1 & -1 & +1 & -1 &-1&+1&+1&-1&- \\
$ E_{1u} $ & +2 & +1 & -1 & -2 & 0 & 0 & -2 & -1 &+1&+2& 0 & 0 & $ (x,y) $ \\
$ E_{2u} $ & +2 & -1 & -1 & +2 & 0 & 0 & -2 & +1 &+1&-2& 0 & 0 & - \\\hline
\end{tabular}
\label{tbl:D6h}
\end{table*}

The Bloch wavefunctions of each band transform according to one of the irreducible representations of $ D_{6h}$.  This ``symmetry label'' of the bands can be determined, e.g., 
by considering which atomic orbitals have a large weight at a given $\mathbf{k}$-space point~\cite{kormanyos2013monolayer}.  We used the corresponding output of the Wien2k code to
obtain the symmetries of nine bands that we will use to set up a  $\mathbf{k}\cdot\mathbf{p}$ model, see Table~\ref{tbl:G-bands-irreps}. Here VB-1, VB-2\dots (CB+1, CB+2\dots)
denotes the first, second band below the VB (above the CB).  
\begin{table}[!htb]
\centering
\caption{Irreducible representations for nine bands at the $ \Gamma $  point. Note, that VB-1 and VB-2 are degenerate, and they are described by the partners of the 
two-dimensional $E_{1g}$ representation. Similarly, CB and CB+1 are degenerate, and  they are described by the partners of the 
two-dimensional $E_{2u}$ representation.}
\begin{tabular}{c c}\hline
Band& Irreducible representation \\\hline
VB-3 & $ B_{1g} $ \\\hline
VB-2 & $ E_{1g} $ \\\hline
VB-1 & $ E_{1g} $ \\\hline
VB & $ A_{2u} $ \\\hline
CB & $ E_{2u} $ \\\hline
CB+1 & $ E_{2u} $ \\\hline
CB+2 & $ A_{1g} $ \\\hline
CB+3 & $ A_{1g} $ \\\hline
CB+4 & $ A_{2u} $ \\\hline
\end{tabular}
\label{tbl:G-bands-irreps}
\end{table}
One can then determine the non-zero matrix elements of the operator  $\mathcal{H}=\frac{1}{2}\left(\hat{p}_{+} q_{}+\hat{p}_{-} q_{+}\right)$, 
where $\hat{p}_{\pm}=\hat{p}_{x}\pm i \hat{p}_{y}$  are momentum operators, using similar arguments as in Ref.~\cite{kormanyos2013monolayer}.
The result is given in Table~\ref{tbl:kp-nine-band-G}. 
\begin{table*}[htb]
 \centering
 \caption{$\mathbf{k}\cdot\mathbf{p}$ matrix elements at   $ \Gamma $ point.}
\begin{tabular}{c c c c c c c c c c}\hline
$H_{k.p}$ &CB+4&CB+3&CB+2&VB-1&VB-2&VB-3&VB&CB&CB+1\\\hline
CB+4 & 0 & 0&0&$ \lambda_{1}q_{-} $&$ \lambda_{2}q_{+} $&0 &0&0&0\\
CB+3 & 0 & 0&0&0&0& 0&0&0&0\\
CB+2 & 0 & 0&0&0&0& 0&0&0&0\\
VB-1& $ \lambda_{1}^{*}q_{+} $ & 0&0&0&0& 0&$ \lambda_{3}q_{+} $&0&$ \lambda_{4}q_{-} $\\
VB-2& $ \lambda_{2}^{*}q_{-} $ & 0&0&0&0& 0&$ \lambda_{5}q_{-} $&$\lambda_{6}q_{+}$&0\\
VB-3& 0 & 0&0&0&0& 0&0&$\lambda_{7}q_{-}$&$\lambda_{8}q_{+}$\\
VB& 0 & 0&0&$ \lambda_{3}^{*}q_{-} $&$ \lambda_{5}^{*}q_{+} $& 0&0&0&0\\
CB & 0 & 0&0&0&$\lambda_{6}^{*}q_{-}$& $\lambda_{7}^{*}q_{+}$&0&0&0\\
CB+1& 0 & 0&0&$\lambda_{4}^{*}q_{+}$&0& $\lambda_{8}^{*}q_{-}$&0&0&0
 \end{tabular}
 \label{tbl:kp-nine-band-G}
 \end{table*}

Some useful relations between the non-zero matrix elements of the $\mathbf{k}\cdot\mathbf{p}$ Hamiltonian 
can be derived by considering the symmetries of the basis functions. 
The degenerate VB-1, VB-2 bands transform as the $E_{1g}$ irreducible representation of $D_{6h}$ at the $\Gamma$ point, see Table~\ref{tbl:G-bands-irreps}. 
Similarly to the degenerate CB, CB+1 bands, the $p_z$ orbitals of the carbon atoms have a large weight in these bands. One can check that the following
linear combinations of the $p_z$ atomic orbitals of the C  atoms transform as the partners of the $E_{1g}$ irreducible
representation:
\begin{subequations}
\begin{align}
&\phi_{3}  = \frac{1}{\sqrt{6}}\left[p_{z}^{1,C}+ \Omega p_z^{2,C}+ \Omega^{2} p_z^{3,C}\right.\nonumber \\
&\qquad \left. - p_z^{4,C}+ \Omega^{4} p_z^{5,C}+ \Omega^{5} p_z^{6,C}\right]
\label{eq:phi3}
\end{align}
\begin{align}
\nonumber
&\phi_{4} = \frac{1}{\sqrt{6}}\left[ p_Z^{1,C}+ \Omega^{5} p_z^{2,C} + \Omega^{4} p_z^{3,C}\right.\nonumber \\
&\qquad \left. -p_z^{4,C}+ \Omega^2 p_z^{5,C} + \Omega P_{6,C}\right]
\label{eq:phi6}
\end{align}
\end{subequations}
where $ \Omega=e^{i \pi/3} $. For simplicity, we denote the Bloch wavefunction based on $\phi_{i}$ by $|\phi_i \rangle$. 
The non-zero matrix elements between the degenerate \{CB,CB+1\} and \{VB-1,VB-2\} bands 
are $\langle \phi_3 |\hat{p}_{-}|\phi_1 \rangle$ and $\langle \phi_4 |\hat{p}_{+}|\phi_2 \rangle$. One can easily check 
that $\left(\langle \phi_4 |\hat{p}_{+}|\phi_2 \rangle\right)^{*}=\langle \phi_3 |\hat{p}_{-}|\phi_1 \rangle$, where $*$ denotes complex conjugation. 

We will denote the  Bloch wavefunction corresponding to VB-3 by $|\phi_{B_{1g}}\rangle$. The non-zero matrix elements between the degenerate \{CB,CB+1\}
and VB-3 are 
$\langle \phi_{B_{1g}} |\hat{p}_{+}|\phi_1 \rangle$ and $\langle \phi_{B_{1g}} |\hat{p}_{-}|\phi_2 \rangle$. Considering the transformation properties of the Bloch wavefunctions 
and of $\hat{p}_{\pm}$ with respect to the $C_{2}^{'}$ rotation, whose axis  is perpendicular to the main, out-of-plane rotation axis, one can show that 
$\langle \phi_{B_{1g}} |\hat{p}_{+}|\phi_1 \rangle=-\langle \phi_{B_{1g}} |\hat{p}_{-}|\phi_2 \rangle$. 
These relations are taken into account in  Table~\ref{tbl:kp-nine-band-G}.

\section{ $ M $ point }
\label{app:M-thirteen-band}

In this Appendix, we explain the most important  details of obtaining the $\mathbf{k}\cdot\mathbf{p}$ Hamiltonian for the $M$ point of the BZ, where 
the relevant point group  is $ D_{2h} $ . The character table for $ D_{2h} $ point group is given  in Table~\ref{tbl:D2h}.  
\begin{table*}[!htb]
\centering
\caption{Character table for $ D_{2h} $  point group.}
\begin{tabular}{c c c c c c c c c c c c}\hline
$D_{2h}$ &$ E$ & $C_{2}(z)$ &$C_{2}(y)$&$C_{2}(x)$ & $ i $& $ \sigma(xy) $&$ \sigma(xz) $& $ \sigma(yz) $& Linear functions,rotations\\\hline
$ A_g $ &  +1 & +1 & +1 & +1 & +1 &+1 &+1 & +1&-  \\
$ B_{1g} $& +1 & +1 & -1 & -1 & +1 & +1 & -1 & -1&$ R_{x} $  \\
$ B_{2g} $& +1 & -1 & +1 & -1 & +1 & -1 & +1 & -1&$ R_{y} $ \\
$ B_{3g} $& +1 & -1 & -1 & +1 & +1 & -1 & -1 & +1&$ R_{z} $  \\
$ A_u $& +1 & +1 & +1 & +1 & -1 & -1 & -1 & -1  & - \\
$ B_{1u} $& +1 & +1 & -1 & -1 & -1 & -1 & +1 & +1&$ x $ \\
$ B_{2u} $& +1 & -1 & +1 & -1 & -1 & +1 & -1 & +1&$ y $ \\
$ B_{3u} $& +1 & -1 & -1 & +1 & -1 & +1 & +1 & -1 &$ z $  \\\hline
\end{tabular}
\label{tbl:D2h}
\end{table*}

Using  information from our DFT calculation, we have assigned an irreducible representations of $ D_{2h}$  to the selected thirteen bands, 
see Table~\ref{tbl:M-bands-irreps}. 
\begin{table}[!htb]
\caption{Irreducible representations for bands at the ${M} $  point. All bands are non-degenerate. }
\begin{tabular}{c c}\hline
Band& Irreducible representation \\\hline
VB-3 & $ B_{3u} $ \\\hline
VB-2 & $ B_{1u} $ \\\hline
VB-1 & $ B_{2g} $ \\\hline
VB & $ B_{3g} $ \\\hline
CB & $ A_{1u} $ \\\hline
CB+1 & $ A_{1g} $ \\\hline
CB+2 & $ B_{2u} $ \\\hline
CB+3 & $ B_{1u} $ \\\hline
CB+4 & $ B_{2u} $ \\\hline
CB+5 & $ A_{1g} $ \\\hline
CB+6 & $ B_{3g} $ \\\hline
CB+7 & $ B_{1u} $ \\\hline
CB+8 & $ B_{3g} $ \\\hline
\end{tabular}
\label{tbl:M-bands-irreps}
\end{table}

Finally, one can set up the  $13$-bands model given in Table~\ref{tbl:kp-13-bands-M}:
\begin{table*}[!htb]
\centering
\caption{$\mathbf{k}\cdot\mathbf{p}$ matrix elements at ${M}$ point}
\scalebox{0.8}{
\begin{tabular}{c c c c c c c c c c c c c c}\hline
$ H_{k.p}$ & VB-3 & VB-2 & VB-1&CB+1&CB+2&CB+3&CB+4&CB+5&CB+6&CB+7&CB+8&VB&CB \\\hline
VB-3 & 0 & 0&0& $\gamma_{1} q_x $&0&0&0&$ \gamma_{2}q_x $&0&0&0&0&0 \\
VB-2 & 0 & 0&$ \gamma_{3}q_x $& 0&0&0&0&0&$ \gamma_{4}q_y $&0&$ \gamma_{5}q_y $&$ \gamma_{6}q_y $&0 \\
VB-1 & 0 & $ \gamma_{3}^{*}q_x $&0& 0&0&$ \gamma_{7}q_x $&0&0&0&$ \gamma_{8}q_x $&0&0&$ \gamma_{9}q_y $ \\
CB+1 & $ \gamma_{1}^{*}q_x $ & 0&0&0&$\gamma_{10}q_y$&0&$\gamma_{11}q_y$&0& 0&0&0&0&0 \\
CB+2 & 0 & 0&0& $ \gamma_{10}^{*}q_y$&0&0&0&0&0&0&0&0&0 \\
CB+3 & 0 & 0&$ \gamma_{7}^{*}q_x $& 0&0&0&0&0&$ \gamma_{12}q_y $&0&$ \gamma_{13}q_y $&$ \gamma_{14}q_y $&0 \\
CB+4 & 0 & 0&0& $\gamma_{11}^{*}q_y$&0&0&0&$\gamma_{15}q_y$&0&0&0&0&0 \\
CB+5 & $ \gamma_{2}^{*}q_x $ & 0&0& 0&0&0&$\gamma_{15}^{*}q_y$&0&0&0&0&0&0 \\
CB+6 & 0 & $ \gamma_{4}^{*}q_y $&0& 0&0&$\gamma_{12}^{*} q_y $&0&0&0&$ \gamma_{16}q_y $&0&0&$ \gamma_{17}q_x $ \\
CB+7 & 0 & 0&$ \gamma_{8}^{*}q_x $& 0&0&0&0&0&$ \gamma_{16}^{*}q_y $&0&$ \gamma_{18}q_y $&$ \gamma_{19}k_y $&0 \\
CB+8 & 0 & $ \gamma_{5}^{*}q_y $&0& 0&0&$\gamma_{13}^{*} q_y $&0&0&0&$ \gamma_{18}^{*}q_y $&0&0&$ \gamma_{20}q_x $ \\
VB & 0 & $ \gamma_{6}^{*}q_y $&0&0&0&$ \gamma_{14}^{*}q_y $&0&0& 0&$ \gamma_{19}^{*}q_y $&0&0&$ \gamma_{21}q_x $ \\
CB & 0 & 0&$ \gamma_{9}^{*}q_y $& 0&0&0&0&0&$ \gamma_{17}^{*}q_x $&0&$\gamma_{20}^{*} q_x $&$\gamma_{21}^{*} q_x $&0 \\\hline
\end{tabular}}
\label{tbl:kp-13-bands-M}
\end{table*}


\bibliographystyle{unsrt}
\bibliography{biblio-c3n}

\end{document}